# A sequential explanatory mixed-methods study on the acceptance of a social robot for EFL speaking practice among Chinese primary school students: Insights from the Computers Are Social Actors (CASA) paradigm

**First Author and Corresponding Author**
Yiran Du
University of Cambridge, Cambridge, United Kingdom
Email: yd392@cam.ac.uk
ORCiD: https://orcid.org/0000-0002-6576-0073

**Second Author**
Jinlong Li
Dalton Xinhua School, Shenzhen, China
Email: 15301115644@163.com

**Third Author**
Huimin He
Xi'an Jiaotong-Liverpool University, Suzhou, China
Email: huimin.he@xjtlu.edu.cn

**Fourth Author**
Chenghao Wang
Xi'an Jiaotong-Liverpool University, Suzhou, China
Email: dancerluo@outlook.com

**Fifth Author**
Bin Zou
Xi'an Jiaotong-Liverpool University, Suzhou, China
Email: bin.zou@xjtlu.edu.cn

**Abstract**
This study investigates Chinese primary school students' acceptance of a social robot for English-as-a-foreign-language (EFL) speaking practice through a sequential explanatory mixed-methods design. Integrating the Technology Acceptance Model (TAM) and the Computers Are Social Actors (CASA) paradigm, the research explores both functional and social factors influencing learners' behavioural intention to use the robot. Quantitative data from 436 students were analysed using structural equation modelling, followed by qualitative interviews with twelve students to interpret the findings. Results show that perceived enjoyment and ease of use are the strongest predictors of acceptance, while social attributes such as warmth, anthropomorphism, and social presence significantly enhance enjoyment. Perceived intelligence affects usefulness but not ease of use. The findings suggest that emotional and social engagement are central to young learners' acceptance of educational robots, highlighting the importance of designing socially intelligent technologies that promote motivation and speaking confidence in EFL learning contexts.

## 1. Introduction
Social robots, physically embodied agents capable of socially meaningful interaction—are increasingly used in education to promote communication and learning (Breazeal et al., 2016; Mavrogiannis et al., 2023). Their physical presence and expressive behaviours enable natural interaction, making them particularly effective for language learning, where social interaction is central to developing speaking competence (Belpaeme et al., 2018; Johal, 2020). Previous studies have shown that robot-assisted learning can enhance learners' motivation, confidence, and performance (Van Den Berghe et al., 2019; Huang & Moore, 2023). However, most research has concentrated on vocabulary and reading development, with limited exploration of speaking practice that requires spontaneous and reciprocal communication (Lampropoulos, 2025; Van Den Berghe et al., 2019).

In China, where English is taught as a foreign language (EFL), authentic opportunities for oral practice are often limited (Lei & Qin, 2022). Social robots may offer a practical means to increase speaking opportunities, yet their effectiveness depends on learners' acceptance. To explain this acceptance, the present study integrates the Technology Acceptance Model (TAM) (Davis & Granić, 2024; Venkatesh, 2000) and the Computers Are Social Actors (CASA) paradigm (Reeves & Nass, 1996; Xu et al., 2022). While TAM focuses on perceptions of usefulness, ease of use, and enjoyment, CASA highlights how social attributes such as social presence, anthropomorphism, warmth, and intelligence influence users' perceptions (Gambino et al., 2020).

Adopting a sequential explanatory mixed-methods design, this study first uses a quantitative phase to investigate what factors, and to what extent, influence Chinese primary school students' acceptance of a social robot for EFL speaking practice. It then follows with a qualitative phase to explore why and how students perceive and respond to the robot in particular ways. Together, these two phases offer a comprehensive understanding of students' acceptance of social robots in primary EFL education. The study addresses the following research questions:

RQ1: What factors, and to what extent, influence Chinese primary school students' acceptance of a social robot for EFL speaking practice?
RQ2: Why and how do students' perceptions of the robot's functional and social attributes shape their behavioural intention to use it?

## 2. Literature Review and Hypothesis Development
### 2.1 Social Robot for Language Learning
A social robot is a physically embodied agent designed to interact and communicate with humans in a socially meaningful way (Breazeal et al., 2016). Unlike virtual agents or chatbots, social robots possess physical presence, expressive behaviours, and social cues that enable natural and engaging interaction (Mavrogiannis et al., 2023). In educational contexts, social robots have been increasingly used as learning companions that can motivate, guide, and support learners through verbal and non-verbal communication (Belpaeme et al., 2018). Their ability to simulate human-like interaction makes them particularly promising for language education, where social engagement and communication play vital roles in learning outcomes (Johal, 2020).

In the field of language learning, social robots have been applied in various educational settings to assist learners in developing linguistic and communicative competence (Van Den Berghe et al., 2019). A growing body of research has shown that robot-assisted language learning can enhance learners' motivation, confidence, and engagement (G. Huang & Moore, 2023). Most existing studies have focused on vocabulary acquisition and reading comprehension, demonstrating that interaction with social robots can facilitate word retention, pronunciation accuracy, and contextual understanding (Van Den Berghe, 2022). However, comparatively few studies have explored how social robots can support speaking practice, which requires more complex and interactive communication (Lampropoulos, 2025; Van Den Berghe et al., 2019).

Research on social robots for speaking practice remains particularly limited in mainland China, where English is taught as a foreign language (EFL) and opportunities for authentic speaking interaction are often constrained (Lei & Qin, 2022). As a result, there is a growing need to investigate how Chinese young learners perceive and accept social robots as speaking partners in EFL contexts. To address this gap, the present study explores Chinese primary school students' acceptance of a social robot designed for EFL speaking practice. By examining their behavioural intention and underlying perceptions, this study aims to provide insights into the potential of social robots to enhance oral English learning in primary education.

### 2.2 The Conceptual Model

The conceptual model of this study integrates the Technology Acceptance Model (TAM) (Davis & Granić, 2024) and the Computers Are Social Actors (CASA) paradigm (Xu et al., 2022) to explain Chinese primary school students' acceptance of a social robot for English-speaking practice (see Figure 1). According to TAM (Venkatesh, 2000), users' behavioural intention to use a technology (BI) is influenced by their perceptions of its usefulness (PU), ease of use (PEOU), and enjoyment (PE). Complementing this framework, the CASA paradigm posits that people apply social rules and expectations to technologies displaying human-like qualities (Gambino et al., 2020). Therefore, four CASA-based constructs, namely perceived social presence (PSP), perceived anthropomorphism (PA), perceived warmth (PW), and perceived intelligence (PI), are included to capture learners' social and emotional responses toward the robot. Together, these TAM and CASA variables form an integrated framework that explains both the functional and social factors influencing learners' acceptance of the social robot. This combined model provides a comprehensive understanding of how young EFL learners evaluate and respond to an embodied, interactive language-learning companion.

**Figure 1. The Conceptual Model**

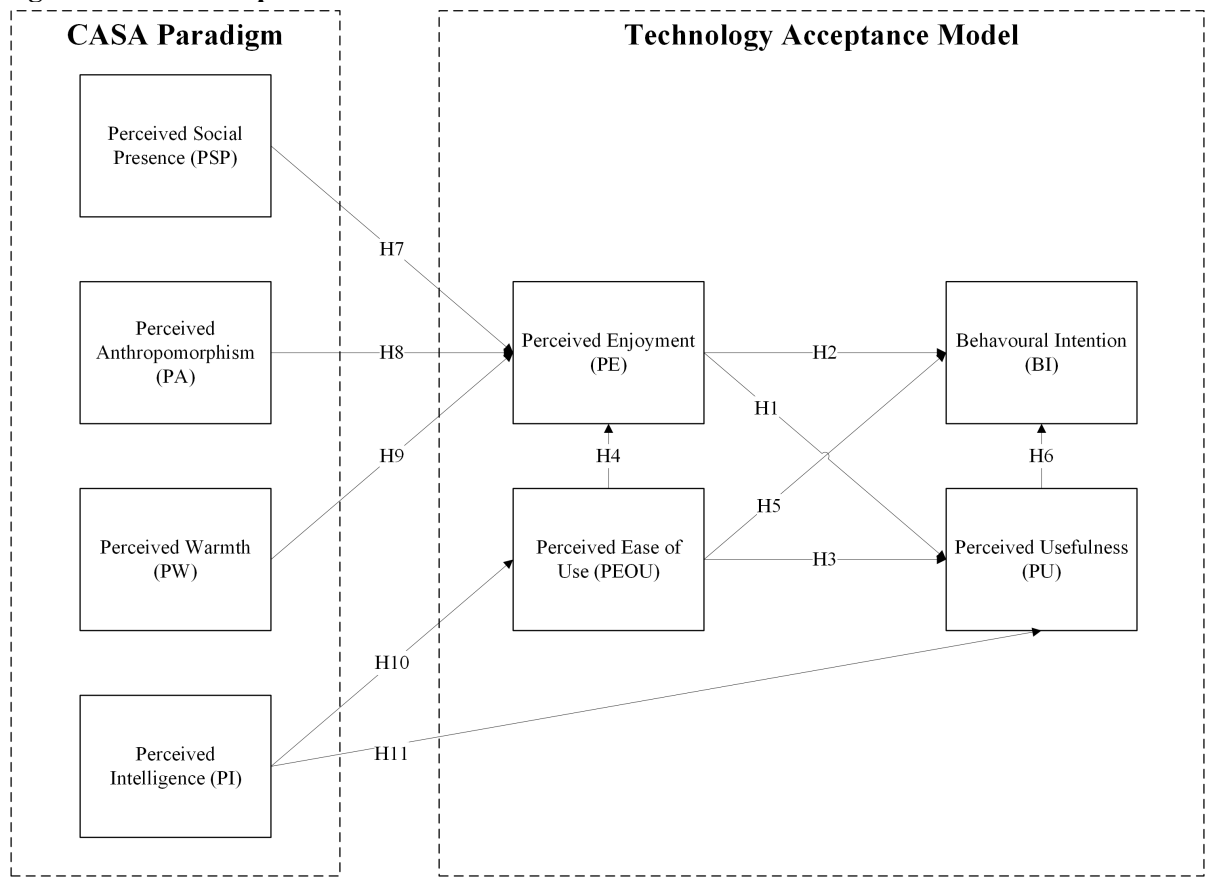

## 2.3 The Roles of Perceived Enjoyment, Ease of Use, and Usefulness

The Technology Acceptance Model (TAM) provides a widely validated framework for explaining how users develop intentions to adopt new technologies. It posits that individuals' behavioural intention (BI) to use a system is primarily determined by their perceived usefulness (PU) and perceived ease of use (PEOU) (Davis, 1989). PU reflects the extent to which users believe that technology enhances their performance, while PEOU refers to the degree to which they find it effortless to operate. Subsequent extensions of TAM have also incorporated perceived enjoyment (PE) as an intrinsic motivational factor, acknowledging that positive emotions play a crucial role in sustaining technology engagement (Teo, 2011; Venkatesh, 2000).

Empirical studies in educational and language learning contexts have provided robust evidence supporting these relationships. PEOU has consistently been shown to influence both PU and PE, as systems that are easier to use are perceived as more beneficial and enjoyable (e.g., An et al., 2024; Cai et al., 2024). Similarly, learners who perceive a technology as useful are more likely to intend to use it for learning purposes (e.g., Hao et al., 2025; Liu et al., 2024). In technology-enhanced and robot-assisted language learning, PE has been found to promote both PU and BI, suggesting that enjoyment not only makes learning more engaging but also enhances perceptions of effectiveness (e.g., F. Huang & Liu, 2024; Lee et al., 2019). Thus, the interrelationships among PE, PEOU, and PU collectively determine learners' willingness to continue using an educational technology.

Building upon TAM and the existing empirical evidence, this study conceptualises the acceptance of a social robot for English-speaking practice among Chinese primary school students as shaped by both extrinsic and intrinsic motivational factors. Learners who find the robot easy to use are expected to perceive it as more enjoyable and useful, which in turn strengthens their intention to continue using it. Similarly, when students experience enjoyment during robot-assisted speaking tasks, they are likely to view the robot as beneficial to their learning and to show stronger adoption intentions. Accordingly, the following hypotheses are proposed:

H1: PE positively influences PU.
H2: PE positively influences BI.
H3: PEOU positively influences PU.
H4: PEOU positively influences PE.
H5: PEOU positively influences BI.
H6: PU positively influences BI.

**2.4 The Roles of Perceived Social Presence, Anthropomorphism, Warmth, and Intelligence**
While the Technology Acceptance Model (TAM) explains learners' cognitive evaluations of technology, it does not fully account for the social and emotional dimensions of interaction that arise when technologies exhibit human-like qualities. The Computers Are Social Actors (CASA) paradigm (Reeves & Nass, 1996) extends this perspective by suggesting that individuals unconsciously apply social rules and expectations to technologies that display social cues such as speech, gesture, and personality. In the context of social robots, this paradigm highlights how users' perceptions of the robot's social attributes can shape their emotional engagement and acceptance (Xu et al., 2022).

Among these social factors, perceived social presence (PSP), perceived anthropomorphism (PA), perceived warmth (PW), and perceived intelligence (PI) have been identified as key constructs influencing user experience. PSP refers to the extent to which a robot is perceived as a real social partner rather than a mechanical device (Oh et al., 2018). High levels of social presence can evoke feelings of connection and involvement, which enhance learners' enjoyment and reduce anxiety during interaction (Hassanein & Head, 2007; Sajjadi et al., 2019). PA captures the degree to which learners attribute human-like characteristics to the robot's appearance and behaviour (Zhou & Zhang, 2024). Robots that exhibit gestures, facial expressions, and conversational cues similar to humans tend to elicit positive emotions and engagement (Kim & Sundar, 2012; Li & Suh, 2022). Similarly, PW, reflecting the robot's friendliness and empathy, has been linked to greater trust and affective bonding, which in turn increase perceived enjoyment in learning interactions (Deng & Yan, 2025; Zheng et al., 2023).

PI, by contrast, reflects learners' perceptions of the robot's cognitive competence, including its ability to understand, respond appropriately, and provide accurate feedback (Tusseyeva et al., 2024). A robot perceived as intelligent is more likely to be seen as easy to use and useful for learning tasks (Song et al., 2024). Prior studies have shown that higher PI enhances both usability and perceived effectiveness in educational technologies (Lampropoulos, 2025; Tusseyeva et al., 2024). Together, these CASA-based constructs explain how social, emotional, and cognitive perceptions shape learners' intrinsic motivation and acceptance beyond the functional determinants proposed by TAM. Therefore, the following hypotheses are proposed:

H7: PSP positively influences PE.
H8: PA positively influences PE.

H9: PW positively influences PE.
H10: PI positively influences PEOU.
H11: PI positively influences PU.

## 3. Methods
### 3.1 Research Design
This study employed a sequential explanatory mixed-methods design to investigate Chinese primary school students' acceptance of a social robot for English-as-a-foreign-language (EFL) speaking practice. The quantitative phase was conducted first, using a structured questionnaire grounded in the integrated Technology Acceptance Model (TAM) and Computers Are Social Actors (CASA) framework. Data from 436 participants were analysed using Structural Equation Modelling (SEM) to examine the hypothesised relationships among perceived functional and social factors influencing behavioural intention. The subsequent qualitative phase aimed to explain and extend the quantitative findings. Semi-structured interviews were carried out with twelve students purposefully selected to represent different acceptance levels, genders, and grade levels. The interview transcripts were analysed thematically to identify recurring patterns, perceptions, and experiences concerning the use of the social robot for EFL speaking practice. The combination of SEM and thematic analysis provided a comprehensive understanding of both the statistical relationships and the underlying reasons shaping students' acceptance of the robot.

### 3.2 Participants
A total of 436 Chinese primary school students participated in the questionnaire phase of the study. All participants were recruited from a private primary school in Shenzhen, China, that follows an international curriculum integrating English as a medium of instruction for several subjects. The students were aged between nine and twelve years ($M = 10.3$, $SD = 0.9$) and were enrolled in Grades 4 to 6. Their English proficiency, assessed according to the Common European Framework of Reference for Languages (CEFR), ranged from A1 to B1, with most achieving the A2 level. All participants had studied English for an average of five to six years and reported varying degrees of experience with robots, primarily through school-based or recreational exposure. Table 1 summarises the characteristics of the questionnaire participants.

**Table 1. Characteristics of Questionnaire Participants (N = 436)**

| Variable | Item | Frequency (n) | Percentage (%) |
| --- | --- | --- | --- |
| Gender | Male | 224 | 51.4 |
|  | Female | 212 | 48.6 |
| Age (years) | 9 | 98 | 22.5 |
|  | 10 | 156 | 35.8 |
|  | 11 | 132 | 30.3 |
|  | 12 | 50 | 11.5 |
| Grade | 4 | 112 | 25.7 |
|  | 5 | 178 | 40.8 |
|  | 6 | 146 | 33.5 |
| English proficiency (CEFR) | A1 | 152 | 34.9 |
|  | A2 | 240 | 55.0 |
|  | B1 | 44 | 10.1 |
| Experience with robots | None | 188 | 43.1 |
|  | Limited | 190 | 43.6 |
|  | Moderate | 58 | 13.3 |

Following the questionnaire, twelve students were selected for semi-structured interviews to explore individual differences in robot acceptance. A stratified purposeful sampling approach (L. Cohen et al., 2018) was applied based on gender (male and female), grade level (4, 5, and 6), and acceptance level, which was determined using each student's mean Behavioural Intention (BI) score. Participants whose mean BI scores were in the upper quartile of the distribution were classified as having high acceptance, while those in the lower quartile were classified as having low acceptance. Within each combination of these three factors (2 × 3 × 2), one participant was selected, resulting in a total of twelve interviewees. To ensure diversity without increasing the sample size, variation in age (9–12 years), English proficiency (A1–B1), and prior experience with robots (none, limited, moderate) was also considered. Table 2 presents the characteristics of the interview participants.

**Table 2. Characteristics of Interview Participants (N = 12)**

| ID | Gender | Grade | Age | Acceptance Level | Proficiency | Experience with Robots |
|---|---|---|---|---|---|---|
| P1 | Male | 4 | 9 | High | A2 | Limited |
| P2 | Male | 4 | 9 | Low | A1 | None |
| P3 | Female | 4 | 10 | High | A2 | Limited |
| P4 | Female | 4 | 9 | Low | A1 | None |
| P5 | Male | 5 | 10 | High | A2 | Limited |
| P6 | Male | 5 | 11 | Low | A1 | None |
| P7 | Female | 5 | 11 | High | A2 | Moderate |
| P8 | Female | 5 | 10 | Low | A1 | Limited |
| P9 | Male | 6 | 11 | High | B1 | Moderate |
| P10 | Male | 6 | 12 | Low | A2 | Limited |
| P11 | Female | 6 | 11 | High | B1 | Moderate |
| P12 | Female | 6 | 12 | Low | A2 | Limited |

### 3.3 The Social Robot

The social robot used in this study, named Lvbao (绿宝 in Chinese, meaning green buddy), was developed as an intelligent speaking companion to assist Chinese primary school students in English-as-a-foreign-language (EFL) speaking practice. Based on the school's mascot and co-designed with student participation, Lvbao embodied a friendly and familiar image that encouraged positive emotional engagement among learners. The robot possessed a real, physical body that was modelled, 3D printed, and assembled using embedded components for speech recognition, voice output, and wireless connectivity, enabling smooth verbal interaction and real-time communication with learners (see Figure 2).

**Figure 2. The 3D model, printing process, and completed Lvbao**

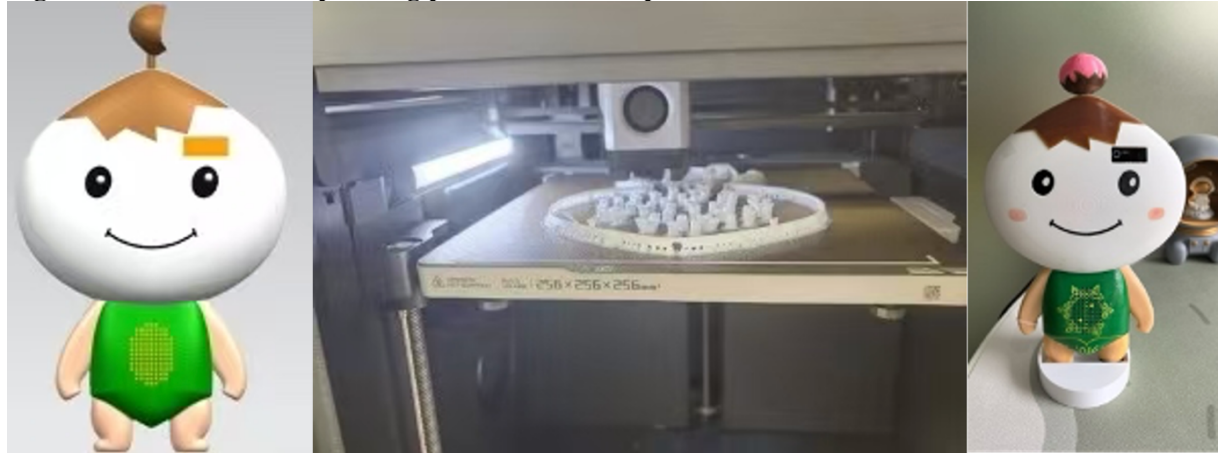

Lvbao was designed to simulate natural spoken communication through its integrated speech recognition, text-to-speech, and cloud-based dialogue system. It could greet learners, introduce speaking activities, ask and answer questions, and offer immediate encouragement or corrective feedback. The robot's functions were supported by an online dialogue platform that managed conversational flow and recorded learners' interactions for later analysis. Through these functions, Lvbao created an engaging and interactive environment that promoted spoken English practice in a playful yet educational way.

The design of Lvbao followed the principles of the Computers Are Social Actors (CASA) paradigm, which suggests that people respond socially to technologies displaying human-like attributes. Four CASA-based factors were deliberately incorporated. Its embodied and responsive nature enhanced perceived social presence, making learners feel they were interacting with a social partner. Its human-like appearance, voice, and gestures supported anthropomorphism, fostering natural and familiar communication. Its friendly tone and supportive feedback strengthened perceived warmth, creating a comfortable learning atmosphere. Finally, its accurate comprehension and appropriate responses reflected perceived intelligence, promoting learner trust and engagement.

By combining technical functionality with social design, Lvbao served not only as a learning tool but also as an emotionally engaging and intelligent partner that encouraged motivation and active participation in EFL speaking activities.

### 3.4 Questionnaires

The quantitative data were obtained through a structured questionnaire developed to examine students' acceptance of the social robot Lvbao for English-speaking practice. The questionnaire included eight constructs adapted from validated instruments grounded in the Technology Acceptance Model (TAM) (Cai et al., 2023; G. Liu & Ma, 2023) and the Computers Are Social Actors (CASA) paradigm (Deng & Yan, 2025; Song et al., 2024; Zheng et al., 2023). Each construct comprised four items measured on a five-point Likert scale ranging from 1 (strongly disagree) to 5 (strongly agree). All items were written in clear, age-appropriate language suitable for primary school students (Y. Du, 2024) (see Table 3 for the full list).

Four constructs from the TAM, namely Perceived Usefulness (PU), Perceived Ease of Use (PEOU), Perceived Enjoyment (PE), and Behavioural Intention to Use (BI), were used to evaluate how students perceived the robot in terms of its effectiveness, usability, enjoyment, and their intention to continue using it in EFL speaking practice. The remaining four constructs from the CASA framework, including Perceived Social Presence (PSP), Perceived Anthropomorphism (PA), Perceived Warmth (PW), and Perceived Intelligence (PI), captured students' social and emotional responses to Lvbao, such as their perceptions of its human-like characteristics, friendliness, and intelligence.

The original English questionnaire was translated into Chinese through a translation and back-translation process to ensure conceptual and linguistic equivalence. A bilingual expert panel reviewed both versions to confirm clarity, cultural appropriateness, and consistency in meaning. A pilot test involving 20 students was conducted to assess the readability and reliability of the items, and minor adjustments were made based on the feedback. The final questionnaire was conducted in Chinese to ensure that all participants could fully understand the items and respond accurately in their native language.

**Table 3. Constructs and Measurement Items**

| Construct and Definition | Measurement Item (English) | Measurement Item (Chinese) |
|---|---|---|
| Perceived Usefulness (PU) | PU1 Using Lvbao improves my English-speaking skills | PU1 使用绿宝能提高我的英语口语能力 |
| Definition: The degree to which using the robot enhances learners' English-speaking performance | PU2 Lvbao helps me practise speaking English more effectively | PU2 绿宝能帮助我更有效地练习英语口语 |
| | PU3 Lvbao makes my speaking practice more productive | PU3 使用绿宝让我的口语练习更有成效 |
| | PU4 Lvbao enhances the quality of my English-speaking activities | PU4 绿宝能提升我英语口语活动的质量 |
| Perceived Ease of Use (PEOU) | PEOU1 Learning to use Lvbao was easy for me | PEOU1 学习使用绿宝对我来说很容易 |
| | PEOU2 I find it simple to interact with Lvbao | PEOU2 与绿宝互动很简单 |
| Definition: The extent to which learners believe that using the robot is free of effort | PEOU3 It is easy to become skilful at using Lvbao | PEOU3 我很容易熟练使用绿宝 |
| | PEOU4 I can use Lvbao without much guidance | PEOU4 我几乎不用别人指导就能使用绿宝 |
| Perceived Enjoyment (PE) | PE1 I enjoy practising English with Lvbao | PE1 我喜欢和绿宝一起练习英语 |
| Definition: The degree to which interacting with the robot is enjoyable beyond functional benefits | PE2 Using Lvbao is fun | PE2 使用绿宝很有趣 |
| | PE3 I feel happy when speaking with Lvbao | PE3 和绿宝说英语时我感到开心 |
| | PE4 I find the experience with Lvbao entertaining | PE4 与绿宝互动的经历让我感到愉快 |
| Perceived Social Presence (PSP) | PSP1 I feel a sense of human contact when interacting with Lvbao | PSP1 与绿宝互动时我感觉像在和人交流 |
| Definition: The feeling of being socially connected and engaged with the robot during interaction | PSP2 Lvbao seems to be aware of me | PSP2 绿宝好像能注意到我 |
| | PSP3 I feel that Lvbao understands my emotions | PSP3 我觉得绿宝能理解我的情绪 |
| | PSP4 I feel personally connected to Lvbao | PSP4 我觉得自己与绿宝有联系 |
| Perceived Anthropomorphism (PA) | PA1 Lvbao behaves like a human | PA1 绿宝的行为像人一样 |
| | PA2 Lvbao's gestures and expressions seem natural | PA2 绿宝的表情和动作很自然 |
| Definition: The extent to which the robot is perceived as human-like in appearance and behaviour | PA3 Lvbao communicates in a human-like way | PA3 绿宝的交流方式像人 |
| | PA4 Lvbao seems to have a personality | PA4 绿宝好像有自己的个性 |
| Perceived Warmth (PW) | PW1 Lvbao seems friendly | PW1 绿宝看起来很友好 |
| | PW2 Lvbao gives me positive and supportive feedback | PW2 绿宝给我积极和支持性的反馈 |
| Definition: The extent to which the robot is perceived as friendly kind and caring | PW3 Lvbao makes me feel comfortable when practising | PW3 绿宝让我在练习时感到放松 |
| | PW4 Lvbao appears kind and caring | PW4 绿宝显得亲切又关心我 |
| Perceived Intelligence (PI) | PI1 Lvbao understands what I say accurately | PI1 绿宝能准确理解我说的话 |
| Definition: The perception that the robot behaves in a competent and intelligent manner | PI2 Lvbao provides appropriate responses to my questions | PI2 绿宝能恰当地回答我的问题 |
| | PI3 Lvbao behaves intelligently during conversation | PI3 绿宝在交流时表现得很聪明 |
| | PI4 Lvbao communicates logically and clearly | PI4 绿宝的交流逻辑清晰 |
| Behavioural Intention (BI) | BI1 I intend to use Lvbao again for English-speaking practice | BI1 我打算再次使用绿宝练习英语口语 |
| Definition: The extent to which learners intend to continue using the robot in the future | BI2 I will recommend Lvbao to my classmates | BI2 我会向同学推荐绿宝 |
| | BI3 I plan to use Lvbao regularly | BI3 我计划经常使用绿宝 |
| | BI4 I would like to continue using Lvbao in future English lessons | BI4 我希望在以后的英语课上继续使用绿宝 |

### 3.5 Semi-Structured Interview

Semi-structured interviews were conducted to gain a deeper understanding of Chinese primary school students' perceptions and experiences when using the social robot Lvbao for English-speaking practice. The interviews focused on how students viewed the robot's usefulness, usability, enjoyability, and social qualities during interaction.

The interview protocol was developed from the integrated Technology Acceptance Model (TAM) and Computers Are Social Actors (CASA) framework. The questions addressed key constructs such as perceived ease of use, usefulness, enjoyment, behavioural intention, social presence, anthropomorphism, warmth, and intelligence. The final protocol comprised thirteen open-ended questions organised into four sections: warm-up and context, TAM-related perceptions, CASA-related perceptions, and integrative reflections (see Appendix A).

Twelve students were selected using stratified purposeful sampling to ensure variation in acceptance level (high and low), gender, and grade level (4–6). Each interview was conducted individually in Chinese to ensure comprehension and natural communication. Interviews took place in a quiet room at the school and lasted approximately 20–30 minutes.

The researcher adopted a friendly and conversational approach appropriate for children aged nine to twelve. Follow-up prompts, such as "Can you tell me more about that?" or "Why do you think so?", were used to encourage elaboration and clarification. All interviews were audio-recorded with parental consent and subsequently transcribed verbatim. The transcripts were translated into English, and the accuracy of translation was checked through back-translation by a bilingual researcher.

### 3.6 Procedure

The study took place between 9 and 24 June 2025 during the English Speaking Challenge organised at the participating school (see Figure 3). In this activity, students individually engaged in an uninterrupted English-speaking interaction with the social robot Lvbao for up to four minutes. Each participant was required to sustain a continuous spoken dialogue with the robot without pauses. The challenge aimed to encourage fluency and persistence in speaking, and students who achieved the longest interaction time or the highest number of dialogue exchanges were recognised as winners.

After completing the speaking challenge, participants filled in a structured questionnaire assessing their perceptions of the robot based on the integrated Technology Acceptance Model (TAM) and Computers Are Social Actors (CASA) framework. Subsequently, semi-structured interviews were conducted with twelve students selected according to acceptance level, gender, and grade to explore their individual experiences and perceptions in greater depth.

Ethical approval for the study was obtained from the research ethics committee of authors' universities. Participation was entirely voluntary, and written informed consent was secured from both students and their parents or legal guardians prior to data collection. All participants were assured of confidentiality and informed that their responses would be used solely for research purposes.

**Figure 3. Students interacting with the social robot Lvbao**

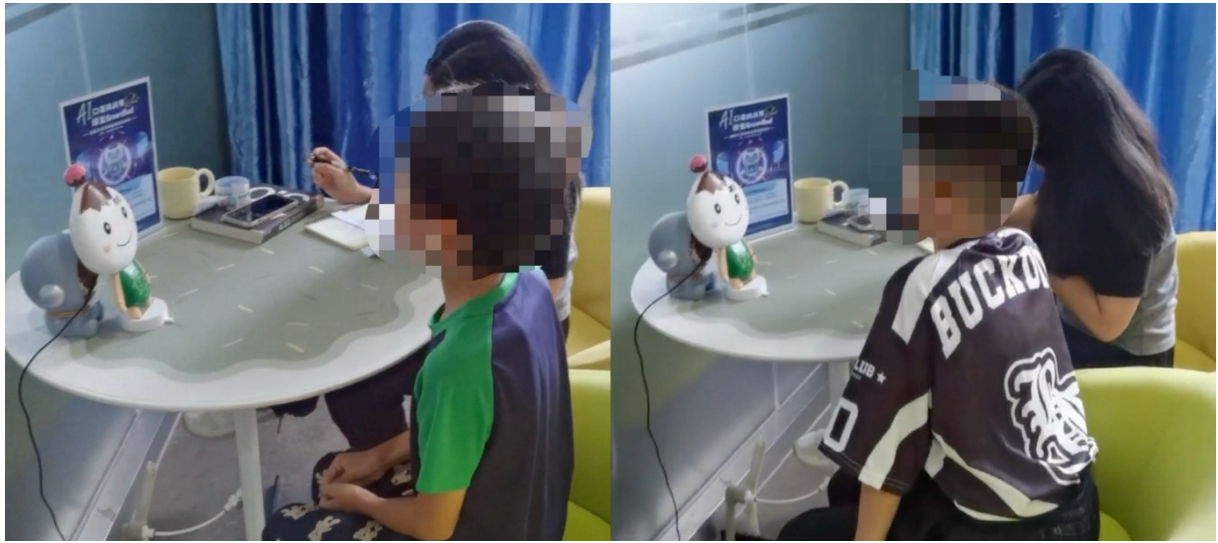

## 3.7 Data Analysis

Quantitative data obtained from the questionnaires were analysed using Structural Equation Modelling (SEM) in R (R Core Team, 2025) with the *lavaan* package (Rosseel, 2012) to examine the hypothesised relationships among constructs from the integrated TAM and CASA framework. Prior to analysis, the dataset was screened for missing values, outliers, and normality. Descriptive statistics, including means, standard deviations, skewness, and kurtosis, were calculated to evaluate participants' perceptions of the social robot. The measurement model was first assessed to determine reliability, convergent validity, and discriminant validity using Cronbach's alpha, composite reliability (CR), and average variance extracted (AVE). Model fit was evaluated through multiple goodness-of-fit indices, including the chi-square to degrees of freedom ratio ($\chi^2$/df), Comparative Fit Index (CFI), Tucker–Lewis Index (TLI), Root Mean Square Error of Approximation (RMSEA), and Standardised Root Mean Square Residual (SRMR). Once the measurement model demonstrated satisfactory psychometric properties, the structural model was estimated to test the significance and strength of the hypothesised paths among the TAM and CASA variables.

The qualitative data from semi-structured interviews were analysed thematically using an inductive–deductive approach to complement and explain the quantitative results. This approach allowed the analysis to be guided by theoretical constructs from the TAM–CASA framework while remaining open to themes emerging directly from students' experiences. All interviews were transcribed verbatim, translated from Chinese into English, and verified for accuracy. Thematic analysis followed Braun and Clarke's (2006) six-phase procedure, and NVivo 14 was used to manage and organise the data. To ensure reliability, two researchers independently coded 25% of the transcripts using R with the *irr* package (Gamer et al., 2019) to calculate inter-coder agreement. Cohen's Kappa reached 0.82, indicating strong reliability (J. Cohen, 1960). Discrepancies were discussed and resolved through consensus before finalising the coding scheme. Integration of quantitative and qualitative findings was carried out during interpretation to provide a comprehensive understanding of students' acceptance of the social robot, linking statistical results with explanatory insights from participants' narratives.

## 4. Results
### 4.1 Quantitative Results
#### 4.1.1 Descriptive Statistics and Normality

Table 4 presents the descriptive statistics for the eight constructs examined in this study, including Perceived Usefulness (PU), Perceived Ease of Use (PEOU), Perceived Enjoyment (PE), Behavioural Intention (BI), Perceived Social Presence (PSP), Perceived Anthropomorphism (PA), Perceived Warmth (PW), and Perceived Intelligence (PI). The mean values ranged from $M = 3.88$ to $M = 4.24$ on a five-point Likert scale, indicating generally positive perceptions of the social robot among the participants. Perceived Enjoyment ($M = 4.24$, $SD = 0.54$) and Perceived Warmth ($M = 4.21$, $SD = 0.53$) showed the highest means, suggesting that students particularly appreciated the enjoyable and friendly nature of the robot during English-speaking practice. Perceived Anthropomorphism ($M = 3.88$, $SD = 0.65$) had the lowest mean, implying that the robot was seen as somewhat less human-like than other aspects such as warmth or intelligence. The standard deviations ranged from $SD = 0.53$ to $SD = 0.65$, indicating moderate dispersion and consistent responses across constructs. All skewness and kurtosis values fell within the acceptable range of −1 to +1, confirming that the data approximated normal distribution and were suitable for subsequent parametric analyses (Kline, 2016). Overall, these findings suggest that Chinese primary school students held favourable perceptions of the social robot's functional and social attributes, supporting its potential as an effective and engaging tool for English-speaking practice.

**Table 4. Descriptive Statistics of the Constructs**

| Construct | *M* | *SD* | Skewness | Kurtosis |
|---|---|---|---|---|
| Perceived Usefulness (PU) | 4.12 | 0.58 | −0.41 | −0.36 |
| Perceived Ease of Use (PEOU) | 4.08 | 0.61 | −0.35 | −0.44 |
| Perceived Enjoyment (PE) | 4.24 | 0.54 | −0.52 | −0.22 |
| Behavioural Intention (BI) | 4.17 | 0.63 | −0.46 | −0.31 |
| Perceived Social Presence (PSP) | 4.09 | 0.57 | −0.39 | −0.37 |
| Perceived Anthropomorphism (PA) | 3.88 | 0.65 | −0.28 | −0.48 |
| Perceived Warmth (PW) | 4.21 | 0.53 | −0.55 | −0.26 |
| Perceived Intelligence (PI) | 4.05 | 0.59 | −0.42 | −0.29 |

### 4.1.2 The Measurement Model

The measurement model was assessed to examine the reliability, convergent validity, and discriminant validity of the eight constructs. As shown in Table 5, all standardised factor loadings exceeded 0.70 ($p < .001$), indicating that each item was a strong indicator of its corresponding latent construct (Kline, 2016). Cronbach's α values ranged from 0.83 to 0.90, and composite reliability (CR) values ranged from 0.87 to 0.93, both surpassing the recommended threshold of 0.70 (Hair, 2019). The average variance extracted (AVE) values ranged from 0.62 to 0.76, exceeding the 0.50 benchmark (Fornell & Larcker, 1981). These results confirm satisfactory internal consistency and convergent validity. Discriminant validity was evaluated using the Fornell–Larcker criterion. As presented in Table 6, the square root of each construct's AVE (diagonal values) was greater than its correlations with other constructs (off-diagonal values), confirming adequate discriminant validity among all latent variables. The overall model fit indices further supported the adequacy of the measurement model (see Table 7). The fit statistics demonstrated a good model fit ($\chi^2/df = 2.41$, CFI = 0.96, TLI = 0.95, RMSEA = 0.06, SRMR = 0.04), all within the recommended thresholds (Hair et al., 2022). Collectively, these results indicate that the measurement model achieved satisfactory reliability, convergent validity, discriminant validity, and overall model fit, providing a sound foundation for testing the structural relationships among constructs.

**Table 5. Reliability and Validity Indicators of the Constructs**

| Construct | Item | Loading | Cronbach's α | CR | AVE |
|---|---|---|---|---|---|
| Perceived Usefulness (PU) | PU1 | 0.81 | 0.87 | 0.90 | 0.69 |
| | PU2 | 0.85 | | | |
| | PU3 | 0.83 | | | |
| | PU4 | 0.82 | | | |
| Perceived Ease of Use (PEOU) | PEOU1 | 0.78 | 0.86 | 0.89 | 0.67 |
| | PEOU2 | 0.84 | | | |
| | PEOU3 | 0.83 | | | |
| | PEOU4 | 0.80 | | | |
| Perceived Enjoyment (PE) | PE1 | 0.84 | 0.88 | 0.91 | 0.72 |
| | PE2 | 0.87 | | | |
| | PE3 | 0.83 | | | |
| | PE4 | 0.85 | | | |
| Behavioural Intention (BI) | BI1 | 0.86 | 0.90 | 0.93 | 0.76 |
| | BI2 | 0.88 | | | |

|  |  | BI3 | 0.87 |  |  |  |
|  |  | BI4 | 0.86 |  |  |  |
| Perceived Social Presence (PSP) |  | PSP1 | 0.79 | 0.85 | 0.88 | 0.65 |
|  |  | PSP2 | 0.82 |  |  |  |
|  |  | PSP3 | 0.81 |  |  |  |
|  |  | PSP4 | 0.80 |  |  |  |
| Perceived Anthropomorphism (PA) |  | PA1 | 0.76 | 0.83 | 0.87 | 0.62 |
|  |  | PA2 | 0.80 |  |  |  |
|  |  | PA3 | 0.81 |  |  |  |
|  |  | PA4 | 0.77 |  |  |  |
| Perceived Warmth (PW) |  | PW1 | 0.85 | 0.88 | 0.91 | 0.71 |
|  |  | PW2 | 0.84 |  |  |  |
|  |  | PW3 | 0.86 |  |  |  |
|  |  | PW4 | 0.83 |  |  |  |
| Perceived Intelligence (PI) |  | PI1 | 0.80 | 0.86 | 0.89 | 0.67 |
|  |  | PI2 | 0.83 |  |  |  |
|  |  | PI3 | 0.82 |  |  |  |
|  |  | PI4 | 0.81 |  |  |  |

**Table 6. Fornell–Larcker Criterion Matrix**

| Construct | PU | PEOU | PE | BI | PSP | PA | PW | PI |
|---|---|---|---|---|---|---|---|---|
| PU | 0.83 |  |  |  |  |  |  |  |
| PEOU | 0.65 | 0.82 |  |  |  |  |  |  |
| PE | 0.60 | 0.62 | 0.85 |  |  |  |  |  |
| BI | 0.67 | 0.61 | 0.66 | 0.87 |  |  |  |  |
| PSP | 0.56 | 0.59 | 0.63 | 0.60 | 0.81 |  |  |  |
| PA | 0.52 | 0.57 | 0.55 | 0.53 | 0.58 | 0.79 |  |  |
| PW | 0.59 | 0.58 | 0.66 | 0.63 | 0.64 | 0.55 | 0.84 |  |
| PI | 0.61 | 0.63 | 0.60 | 0.62 | 0.59 | 0.56 | 0.61 | 0.82 |

Notes. Bolded diagonal values represent the square root of AVE for each construct. Off-diagonal values represent the inter-construct correlations.

**Table 7. Model Fit Indices for the Measurement and Structural Models**

| Fit Index | Threshold | Measurement Model | Structural Model |
|---|---|---|---|
| $\chi^2/df$ | < 3.00 | 2.41 | 2.58 |
| CFI | ⩾ 0.90 | 0.96 | 0.95 |
| TLI | ⩾ 0.90 | 0.95 | 0.94 |
| RMSEA | ⩽ 0.08 | 0.06 | 0.06 |
| SRMR | ⩽ 0.08 | 0.04 | 0.05 |

### 4.1.3 The Structural Model

The structural model was tested to examine the hypothesised relationships among the constructs derived from the integrated Technology Acceptance Model (TAM) and Computers Are Social Actors (CASA) framework. As shown in Table 7, the model demonstrated an acceptable overall fit to the data ($\chi^2/df$ = 2.58, CFI = 0.95, TLI = 0.94, RMSEA = 0.06, SRMR = 0.05), all within the recommended thresholds (Hair et al., 2022). These indices indicate that the structural model adequately represented the empirical data and provided a sound basis for hypothesis testing.

Table 8 and Figure 4 present the standardised path coefficients and the corresponding significance levels for the hypothesised relationships. Of the eleven proposed hypotheses, nine were supported. Within the TAM-related paths, perceived enjoyment (PE) significantly influenced perceived usefulness (PU) ($\beta$ = 0.32, $p$ < .001) and behavioural intention (BI) ($\beta$ = 0.27, $p$ < .001). Perceived ease of use (PEOU) exerted strong positive effects on perceived usefulness ($\beta$ = 0.34, $p$ < .001) and perceived enjoyment ($\beta$ = 0.29, $p$ < .001), as well as a weaker but significant effect on behavioural intention ($\beta$ = 0.10, $p$ < .05). However, the direct effect of perceived usefulness on behavioural intention (H6) was not significant ($\beta = 0.08$, $p = .16$).

Regarding the CASA-related constructs, perceived social presence (PSP), perceived anthropomorphism (PA), and perceived warmth (PW) all had significant positive effects on perceived enjoyment ($\beta = 0.21$, $p < .001$; $\beta = 0.15$, $p < .01$; $\beta = 0.25$, $p < .001$, respectively). Perceived intelligence (PI) significantly predicted perceived usefulness ($\beta = 0.26$, $p < .001$) but did not have a significant effect on perceived ease of use ($\beta = 0.09$, $p = .13$).

**Table 8. Structural Model Results and Hypothesis Testing**

| Hypothesis | Path | β | SE | z | Result |
|---|---|---|---|---|---|
| H1 | PE → PU | 0.32 | 0.05 | 6.24*** | Supported |
| H2 | PE → BI | 0.27 | 0.05 | 5.18*** | Supported |
| H3 | PEOU → PU | 0.34 | 0.05 | 6.73*** | Supported |
| H4 | PEOU → PE | 0.29 | 0.05 | 5.92*** | Supported |
| H5 | PEOU → BI | 0.10 | 0.05 | 2.08* | Supported |
| H6 | PU → BI | 0.08 | 0.06 | 1.42 | Not supported |
| H7 | PSP → PE | 0.21 | 0.05 | 4.61*** | Supported |
| H8 | PA → PE | 0.15 | 0.05 | 3.29** | Supported |
| H9 | PW → PE | 0.25 | 0.05 | 5.12*** | Supported |
| H10 | PI → PEOU | 0.09 | 0.06 | 1.53 | Not supported |
| H11 | PI → PU | 0.26 | 0.05 | 5.44*** | Supported |

**Note.** Statistical significance is denoted as *** $p < .001$, ** $p < .01$, * $p < .05$.

**Figure 4. Structural Model Results and Hypothesis Testing**

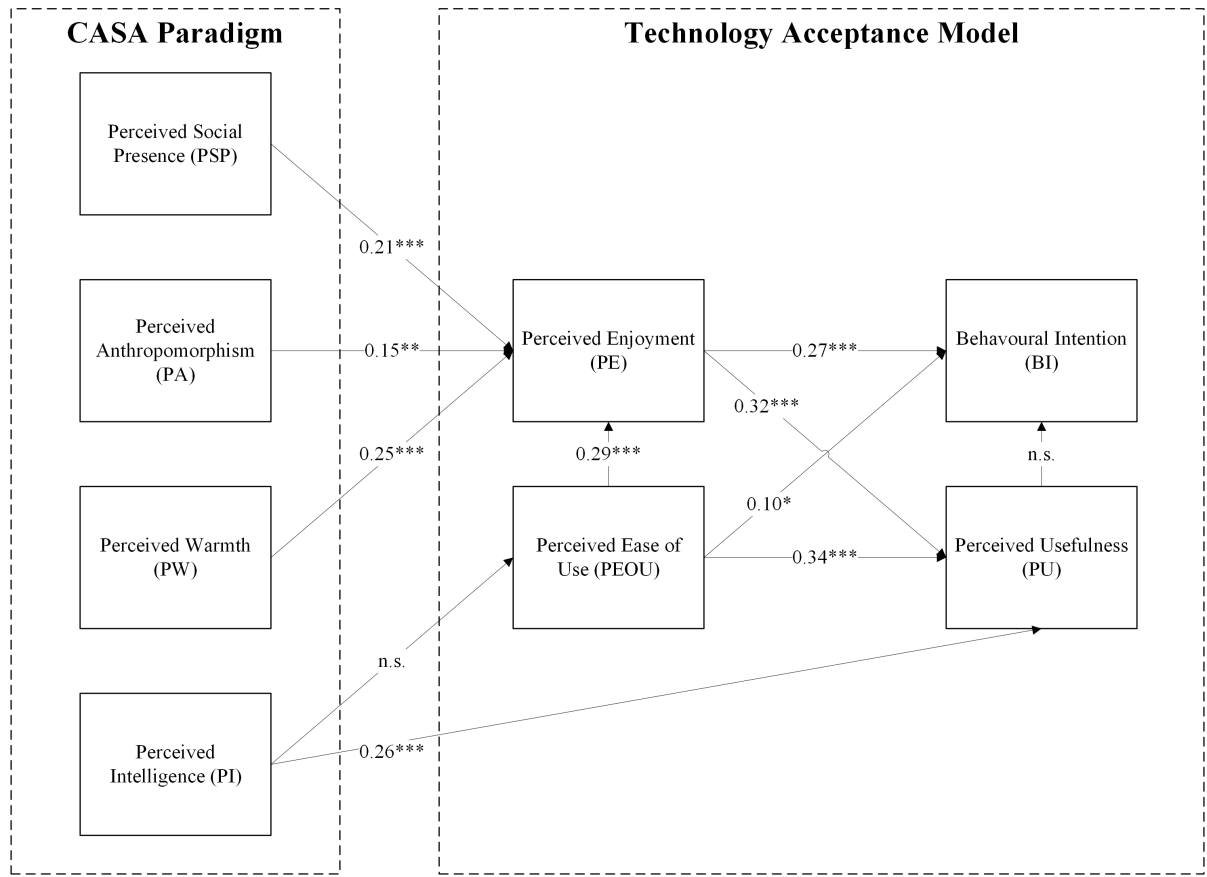

Note. Statistical significance is denoted as *** for $p < 0.001$, ** for $p < 0.01$, * for $p < 0.05$, and n.s. for not significant.

### 4.2 Qualitative Results

The thematic analysis identified three interconnected themes that explain students' acceptance of Lvbao in relation to the study's hypotheses: functional ease and motivational engagement (H1–H6), social presence and emotional connection (H7–H9), and cognitive intelligence and perceived competence (H10–H11).

#### 4.2.1 Functional Ease and Motivational Engagement (H1–H6)

Students frequently mentioned that Lvbao was easy to use, which helped them enjoy speaking with it and find it useful. This supports the relationships between perceived ease of use, enjoyment, and usefulness (H3–H4). P3 explained, "It's easy to talk to Lvbao. I just say something and it answers straight away." P9 added, "I don't have to do anything special. It just understands me." The effortless interaction encouraged confidence and willingness to speak.

Students also linked enjoyment to both usefulness and their intention to continue using the robot (H1–H2). Many said they liked Lvbao because it made English practice enjoyable. P5 stated, "It's fun, like talking to a friend." P11 said, "It makes English more interesting, so I want to talk more." These comments show that the fun and relaxed atmosphere made them believe that Lvbao was helpful for practising English and worth using again.

Ease of use was further connected with intention to use (H5). When students found the robot simple to interact with, they were more likely to engage actively. P4 noted, "It's easy to start talking, so I don't feel nervous." Similarly, P8 said, "If it were hard to use, I would stop, but it's easy." These views indicate that usability encouraged repeated participation and reduced hesitation in speaking English.

Finally, several students valued Lvbao more for enjoyment than for its direct learning benefits, which helps explain why usefulness did not strongly influence their intention to use it (H6). P2 said, "I like using it because it's funny to talk to." P7 added, "Sometimes it asks the same question, but it still makes me laugh." Such remarks suggest that students' motivation was driven mainly by enjoyment rather than perceived improvement, clarifying the weaker relationship between usefulness and intention.

#### 4.2.2 Social Presence and Emotional Connection (H7–H9)

Students often described Lvbao as friendly and attentive, showing how its social presence enhanced their enjoyment of speaking (H7). P9 said, "It looks at me and listens, so I feel it's really talking to me." P11 mentioned, "When it nods, I feel it understands me." These descriptions indicate that Lvbao's responsive behaviours created a sense of being in a real conversation, which made the activity more engaging.

Students also commented on the robot's human-like qualities, linking anthropomorphism with enjoyment (H8). P3 noted, "It moves and blinks like a person." P6 said, "Its voice sounds like a person, not a machine." These features made students feel comfortable speaking to it as if to a partner. However, some recognised its limits. P8 explained, "Sometimes it doesn't understand what I say, so I remember it's just a robot." This suggests that while anthropomorphism encouraged social connection, students remained aware of its artificial nature.

Warmth was another factor contributing to positive feelings (H9). Students appreciated Lvbao's kind tone and encouragement. P5 said, "When it says 'Good job', I feel happy." P4 added, "It talks nicely, so I'm not scared to answer." Such comments show that friendliness and encouragement helped students relax and enjoy the conversation.

#### 4.2.3 Cognitive Intelligence and Perceived Competence (H10–H11)

Students frequently described Lvbao as smart and capable, showing how perceptions of intelligence contributed to its usefulness (H11). P5 said, "It knows when I say something wrong and helps me fix it." P9 added, "It understands what I say and gives a good answer." Many students associated this kind of intelligent response with learning support, saying it made Lvbao a helpful speaking partner.

However, students did not always connect intelligence with ease of use (H10). Several explained that even though Lvbao seemed clever, speaking to it still required effort. P10 said, "It's smart, but sometimes I don't know how to make it answer." P4 shared, "It understands me, but I have to think carefully about what to say." These comments suggest that students viewed intelligence as the robot's thinking ability rather than its usability. For them, a "smart" robot was one that knew things or gave good answers, not necessarily one that was easy to use.

## 5. Discussion

### 5.1 The Impact of Perceived Enjoyment, Ease of Use, and Usefulness

The findings reaffirm the Technology Acceptance Model's (TAM) emphasis on enjoyment and ease of use as key predictors of technology acceptance (Davis, 1989; Davis & Granić, 2024). Perceived enjoyment had the strongest effect on both usefulness and behavioural intention, indicating that children's engagement with the social robot was primarily driven by intrinsic motivation. This aligns with previous research showing that enjoyment fosters both perceived effectiveness and sustained technology use in learning (C. Du et al., 2025; Wang et al., 2024). Students described the interaction with Lvbao as enjoyable and playful, suggesting that emotional satisfaction motivated them more than functional evaluation. Ease of use also strongly influenced enjoyment and usefulness, showing that intuitive interaction was central to children's acceptance. When students felt that Lvbao understood them easily and responded quickly, they were more confident and engaged, consistent with findings that usability promotes flow and reduces cognitive load (Zou, Du, et al., 2023; Zou, Lyu, et al., 2023).

However, perceived usefulness did not significantly predict behavioural intention. This result diverges from typical TAM findings but could be explained by the participants' age. As primary school students, they were less concerned with learning efficiency and more focused on enjoyment and social interaction. Similar research has shown that young learners value fun and emotional connection over measurable academic benefit (Limone & Toto, 2021; Plowman & McPake, 2013). For them, the robot's appeal could lay in its friendly and engaging behaviour rather than in its perceived contribution to language improvement. Thus, the motivational impact of enjoyment and ease of use outweighed the cognitive assessment of usefulness, highlighting the child-centred nature of technology acceptance in this context.

### 5.2 The Impact of Perceived Social Presence, Anthropomorphism, Warmth, and Intelligence

The social and emotional attributes of the robot were central to students' acceptance, consistent with the Computers Are Social Actors (CASA) paradigm (Reeves & Nass, 1996; Xu et al., 2022). Perceived social presence, anthropomorphism, and warmth all significantly increased enjoyment, indicating that learners responded to Lvbao as a social companion. When the robot showed attentive listening and friendly gestures, students felt more relaxed and engaged, supporting earlier evidence that social presence enhances motivation and reduces anxiety in learning (Chen, 2022). Warmth and anthropomorphism were particularly effective in sustaining interest. Lvbao's supportive tone and expressive behaviour created a reassuring environment, consistent with findings that friendliness and human-like cues foster emotional connection and trust (Deng & Yan, 2025; Kim & Sundar, 2012; Zheng et al., 2023). These social attributes appear to be especially meaningful for children, who are naturally responsive to empathy and playfulness in interaction.

Perceived intelligence, however, only influenced usefulness, not ease of use. Students associated intelligence with the robot's ability to understand and respond accurately, rather than how simple it was to operate. This non-significant path may be explained by the participants' age: as children, they are more attuned to the robot's behaviour and personality than to its technical efficiency. Similar findings have been reported in studies where younger users valued social appeal over functional precision (Harverson et al., 2025; Undheim, 2022). Overall, the results suggest that for child users, emotional and social engagement outweigh functional considerations, and perceptions of intelligence are interpreted in social rather than technical terms.

## 5.3 Theoretical and Practical Implications

Theoretically, this study extends the Technology Acceptance Model (TAM) and the Computers Are Social Actors (CASA) paradigm by integrating functional, emotional, and social determinants into a unified framework for understanding children's acceptance of educational robots. By demonstrating that perceived enjoyment and social attributes such as warmth, anthropomorphism, and social presence exert stronger influences than perceived usefulness, the study highlights the predominance of affective and relational factors in young learners' technology acceptance. This finding refines TAM's applicability to child users, suggesting that motivation and enjoyment serve as primary drivers when learning technologies are designed with embodied and interactive features. Moreover, the integration of CASA constructs into TAM contributes to a more comprehensive theoretical understanding of human–robot interaction in education, revealing how children interpret social cues and attribute agency to artificial partners in ways that differ from adult users.

Practically, the findings offer valuable insights for educators, designers, and policymakers involved in robot-assisted language learning. For designers, the results suggest that enhancing a robot's social and emotional expressiveness, such as through friendly feedback, responsive gestures, and human-like voice, may be more effective in promoting engagement than focusing solely on cognitive or instructional capabilities (Y. Du, 2025). Educators can use social robots like Lvbao to create low-anxiety, interactive speaking environments that stimulate motivation and confidence (Wang et al., 2025), particularly for young EFL learners who lack authentic oral practice opportunities. Policymakers and curriculum developers should recognise the pedagogical potential of socially intelligent robots and integrate them thoughtfully into early language education, ensuring that their use complements rather than replaces human interaction. Collectively, these implications underline the importance of designing educational technologies that balance functional utility with emotional resonance to maximise learning impact and long-term acceptance among children.

## 5.4 Limitations and Future Research

Despite offering meaningful insights, this study has several limitations that should be acknowledged. First, the data were collected from a single private primary school in Shenzhen, which may limit the generalisability of the findings to broader contexts. Students in international or urban schools may have higher exposure to technology and English-speaking environments than those in rural or public schools. Future research could include participants from diverse educational settings and regions to capture a wider range of socio-economic and linguistic backgrounds (Y. Du et al., 2025).

Second, the study adopted a cross-sectional design, which restricts the ability to infer long-term behavioural patterns. Longitudinal or experimental studies could provide stronger evidence of how students' perceptions and acceptance evolve with sustained interaction and increasing familiarity with social robots.

Third, while this study integrated quantitative and qualitative data, the qualitative phase involved a small number of interview participants. Expanding the qualitative sample or incorporating classroom observations could yield richer insights into how students' social and emotional engagement unfolds in authentic learning situations.

Finally, the study focused primarily on students' perspectives, without considering teachers' or parents' attitudes toward robot-assisted learning, which may influence students' acceptance and continued use. Future research should therefore adopt a multi-stakeholder approach, examining how teacher guidance, parental support, and classroom dynamics interact with children's perceptions of educational robots.

## 6. Conclusion

In conclusion, this study provides a comprehensive understanding of Chinese primary school students' acceptance of a social robot for EFL speaking practice by integrating the Technology Acceptance Model (TAM) and the Computers Are Social Actors (CASA) paradigm. Quantitative and qualitative findings revealed that perceived enjoyment and ease of use were the strongest predictors of acceptance, while social attributes such as warmth, anthropomorphism, and social presence significantly enhanced students' enjoyment and engagement. Perceived intelligence influenced usefulness but not usability, reflecting children's social interpretation of "smartness." Overall, the results highlight that young learners' acceptance of educational robots is primarily shaped by emotional and social experiences rather than purely functional evaluations. The study advances theoretical understanding by extending TAM to child–robot interaction and offers practical implications for designing socially and emotionally responsive robots that foster motivation, confidence, and oral communication in primary EFL education.

**Appendices**
**Appendix A. Semi-Structured Interview Protocol**

The semi-structured interview aimed to explain and expand upon the quantitative findings by exploring Chinese primary school students' perceptions and experiences of interacting with the social robot Lvbao for English-as-a-foreign-language (EFL) speaking practice. The questions were guided by the integrated Technology Acceptance Model (TAM) and Computers Are Social Actors (CASA) framework. Interviews were conducted in Chinese to ensure clarity and natural expression. Additional prompts (for example, "Can you tell me more about that?" or "Why do you think so?") were used as needed to encourage elaboration.

**Section A. Warm-Up and Context**
1. Can you tell me about your experience of using Lvbao during English speaking practice? What activities did you do with it, and how did you feel?
2. Compared with practising English with a teacher, classmates, or computer software, how did using Lvbao feel different or similar?

**Section B. Questions Related to the Technology Acceptance Model (TAM)**
3. How easy or difficult was it to use Lvbao? Were there any parts that felt confusing or simple to you?
4. Did you enjoy practising English with Lvbao? Which parts were the most fun or frustrating?
5. Do you think Lvbao helped you improve your English speaking or confidence? In what ways?
6. Would you like to continue using Lvbao in the future or recommend it to others? What might make you want to use it more often?

**Section C. Questions Related to the Computers Are Social Actors (CASA) Constructs**
7. When talking with Lvbao, did it feel like a real social partner or more like a machine? How did that affect your speaking experience?
8. What did you think about Lvbao's appearance, voice, and movements? Did these make it feel human-like or friendly to you?
9. How did Lvbao make you feel when it responded, gave feedback, or encouraged you? Did it seem warm or caring?
10. Did you feel that Lvbao understood what you said and gave appropriate responses? How did its intelligence (or mistakes) affect your learning experience?

**Section D. Integrative and Reflective Questions**
11. Overall, what were the best things and main problems about using Lvbao for English speaking practice?
12. If you could design your own speaking robot, what would you change or keep the same?
13. How would you describe Lvbao in three words?